# Graph-based Multi-View Fusion and Local Adaptation: Mitigating Within-Household Confusability for Speaker Identification

*Long Chen, Yixiong Meng, Venkatesh Ravichandran, Andreas Stolcke*

Amazon Alexa AI, USA

`{longchn, myixiong, veravic, stolcke}@amazon.com`

## Abstract

Speaker identification (SID) in the household scenario (e.g., for smart speakers) is an important but challenging problem due to limited number of labeled (enrollment) utterances, confusable voices, and demographic imbalances. Conventional speaker recognition systems generalize from a large random sample of speakers, causing the recognition to underperform for households drawn from specific cohorts or otherwise exhibiting high confusability. In this work, we propose a graph-based semi-supervised learning approach to improve household-level SID accuracy and robustness with locally adapted graph normalization and multi-signal fusion with multi-view graphs. Unlike other work on household SID, fairness, and signal fusion, this work focuses on speaker label inference (scoring) and provides a simple solution to realize household-specific adaptation and multi-signal fusion without tuning the embeddings or training a fusion network. Experiments on the VoxCeleb dataset demonstrate that our approach consistently improves the performance across households with different customer cohorts and degrees of confusability.

## 1. Introduction

AI smart speakers (such as Amazon Echo and Google Home) are used more and more widely, allowing customers to access a large variety of services and experiences by voice. Devices are typically used by multiple speakers within a household. Thus, speaker identification (SID) in a household scenario is the key to enable personized experiences, such as playing a user's preferred music, customizing calendars, as well as authentication for secured services, like shopping.

Compared to a conventional SID task on a fixed set of random speakers, SID for household is more challenging in several aspects. First, SID in the household scenario typically has to be based on only a few enrollment utterances. While a much larger set of unlabeled data is available for each household, annotating the identities of unfamiliar speakers from audio data alone is challenging, with quality, scalability and privacy concerns. Therefore, certain standard SID approaches are not applicable: fully supervised training on the full set of known speakers, typically employing a full-set classifier and requiring pre-defined classes. Second, the speaker embeddings are typically trained with large speaker datasets such as VoxCeleb [1][2] to optimize an averaged speaker verification loss. This training process tends to optimize models for majority speaker demographics and overlooks possibly underrepresented groups (such as nonnative or regional accents, or some age groups or genders), results in lower recognition performance for minority groups and unreliable access to personalized services on smart speakers. This is a general challenge for all AI based on deep neural networks; it has been discussed for face recognition [3], recommender systems [4] and speech applications [5]–[7]. Third, speakers in the same household often share similar voice characteristics and acoustic conditions [8], making SID intrinsically harder than for random speakers. Thus, speaker embeddings, as well as the inference methods trained on a large set of random speakers are not necessarily optimal for household SID. Moreover, SID models as well as hyperparameters optimized for some households may not work for others, considering the diversity of characteristics and different confusability across households.

To address these challenges, we propose a graph-based semi-supervised learning (graph-SSL) method based on label propagation for SID in household scenarios, focusing especially on model fairness across demographic patterns and voice similarities of households. Our previous work [9] has demonstrated the effectiveness of graph label propagation for improving SID with unlabeled data. While the earlier work focused on SSL in households with randomly sampled speakers and universal graph hyperparameters, in this work we propose a graph normalization method that locally adapts to voice similarity in a given household, based on a K-nearest neighbors algorithm. By doing so, we overcome the majoritarian bias and provide more performance fairness across households with different customer cohorts. Moreover, to further improve SID accuracy, we perform multi-signal fusion on the speaker similarity graph (multi-view graph-SSL). Specifically, we investigate the fusion of voice signal, face signal, and session metainformation, with two fusion methods: either edge-level fusion with max-pooling [10] or graph-level fusion with the power mean Laplacian [11]. To the best of our knowledge, this is the first work realizing adaptation and multi-signal fusion with graph-SSL for speaker identification.

In contrast to other recently proposed household-level SID approaches [8][12] or fairness-driven speaker recognition models [5], which focus on generating adapted embeddings or household-specific models, our approach focuses on speaker label inference (scoring) given existing embedding extractors, giving a simple, low-cost solution for better model performance without tuning the embeddings. Moreover, unlike other work [13][14] on signal (mostly audio-visual) fusion for supervised speaker recognition training, our work leverages unlabeled data with graph-SSL, thereby addressing the more realistic household scenario for smart speakers.

## 2. Methods

### 2.1. Problem formulation

Let us assume a household with $C$ speakers (classes). Let $X = \{X^{(1)}...X^{(V)}\}$ denote a multi-view representation of the utterances with $V$ views, where $X^{(v)} = \{x_1^{(v)} ... x_l^{(v)}, x_{l+1}^{(v)} ... x_{l+u}^{(v)}\} \subset R^{D^{(v)}}$ are the embeddings of the utterances in the $v$-th view. Let the first $l$ utterances be labeled samples with $Y_L = \{y_1 ... y_l\} \subset \{1 ... C\}$ being the speaker labels. Let the remaining $u$ utterances with indices from $l+1$ to $l+u$ be the unlabeled utterances, where $Y_U = \{y_{l+1} ... y_{l+u}\} \subset \{1 ... C\}$ are the unknown speaker labels. The problem is to predict $Y_U$ from $X$ and $Y_L$.

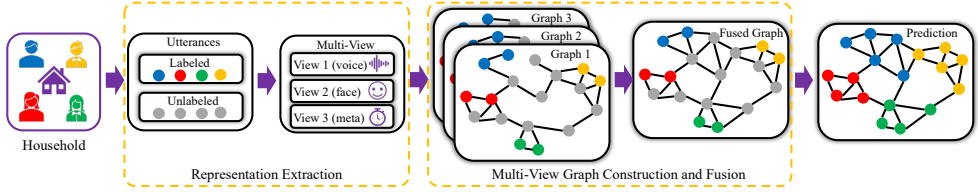

Figure 1: *Overview of the multi-view graph-SSL*

### 2.2. Single-view graph construction

We create a fully connected graph for each household where each node represents an utterance and each weighted edge connecting two nodes quantifies the similarity between the two utterances by its edge weight. The number of nodes in a graph equals the number of (labeled or unlabeled) utterances in the household. There are various ways to measure the similarities with given utterance representations. Here we use Euclidean distance between utterance embeddings to define the edge weight between two nodes $i, j$. For the $v$-th view:

$$W_{ij}^{(v)} = \exp\left(-\frac{dist(x_i^{(v)}, x_j^{(v)})^2}{(\sigma_{ij}^{(v)})^2}\right) \quad (1)$$

where $dist(x, y)$ is the Euclidean distance, $\sigma_{ij}^{(v)}$ is a temperature-like scaling hyperparameter of the model and $W^{(v)}$ is the affinity matrix of edge weights.

### 2.3. Locally adapted graph normalization

In most previous work [15]–[17] including ours [9], a universal scaling factor $\sigma^{(v)}$ is used:

$$\sigma_{ij}^{(v)} = \sigma^{(v)} \quad (2)$$

However, this assumption could introduce majoritarian bias, considering the different degrees of voice similarity among speakers in different households. Thus, we locally adapt the scaling factor based on the mean distance to the K-nearest neighbors [18] to normalize the affinity matrix:

$$\sigma_{ij}^{(v)} = s \cdot mean(knnd(x_i^{(v)}) \cup knnd(x_j^{(v)})) \quad (3)$$

where $knnd(x)$ is the set of distances to the K-nearest neighbors of sample $x$, and $s$ is a hyperparameter of the model. Both $K$ and $s$ are universal across different households, where $K$ controls the locality of the scaling and $s$ the overall sharpness of the kernel function.

### 2.4. Multi-view fusion

We investigate two ways to fuse multiple views of the utterance data. The first, **EdgePool**, uses edge-level max-pooling of weights across different views [10]:

$$W = \max([W^{(1)}, ..., W^{(V)}], axis = 0) \quad (4)$$

where $W$ is the final affinity matrix after fusion.

The second approach, **PML**, involves graph-level fusion with the power mean Laplacian [11]:

$$L_{sym} = \left(\frac{1}{V}\sum_{v=1}^{V}(L_{sym}^{(v)})^p\right)^{1/p} \quad (5)$$

where $L_{sym}^{(v)}$ is the symmetric normalized Laplacian matrix of the graph in the $v$-th view: $L_{sym}^{(v)} = I - D^{(v)-\frac{1}{2}}W^{(v)}D^{(v)-\frac{1}{2}}$. $D^{(v)}$ is the degree diagonal matrix with $D_{ii}^{(v)} = \sum_{j=1}^{l+u}W_{ij}^{(v)}$ and $p$ is a real constant defining the power. $L_{sym}$ is the final symmetric normalized Laplacian matrix after fusion.

### 2.5. Label propagation

Label propagation (LP) [15]–[17] is a simple but effective transductive learning approach widely used for graph-SSL. The basic idea of LP is to find a joint labeling of all nodes in the graph such that (1) the labels that are given a priori are not changed too much and (2) the labeling is smooth over the graph. Here we employ the following objective function:

$$\operatorname{argmin}_{\mathbf{f}} (\|\mathbf{f} - \mathbf{Y}\|_2^2 + \lambda\, trace(\mathbf{f}^T L_{sym}\mathbf{f})) \quad (6)$$

where $\mathbf{Y}$ is the input matrix representation of known labels, $\mathbf{f}$ is the labeling solution, and $\lambda$ is a regularization hyperparameter. $L_{sym}$ is the symmetric normalized Laplacian matrix of the graph as mentioned above. The first term of the objective function is the supervised loss and the second term is the graph-regularization term that ensures smoothness, i.e., label consistence of nearby samples. To solve this objective function for each household, we employ the iterative algorithm introduced by Zhou et al. [16]. The method aims to spread every sample's label information through the graph until achieving global convergence. We add a class normalization operation to minimize the influence of imbalance in the labels/pseudo-labels [19], as shown in Algorithm 1.

| Algorithm 1: Label Propagation with Normalization |
|---|
| Compute the affinity matrix $W$ if $i \neq j$ & $W_{ii} = 0$; or directly use $L_{sym}$ if is input |
| Compute matrix $S = D^{-1/2}WD^{-1/2}$ or $S = I - L_{sym}$ |
| Initialize $\hat{Y}^{(0)}$: $\hat{Y}_{ij}^{(0)} = \begin{cases} 1 \text{ if } i \leq l, \ x_i \text{ is labeled as } j \\ 0 \text{ otherwise} \end{cases}$ |
| Normalize $\hat{Y}^{(0)}$: $\hat{Y}_{ij}^{(0)} = \hat{Y}_{ij}^{(0)} / \sum_k \hat{Y}_{kj}^{(0)}$ |
| Choose a parameter $\alpha \in (0,1)$ |
| Iterate $\hat{Y}^{(t+1)} = \alpha S \hat{Y}^{(t)} + (1-\alpha)\hat{Y}^{(0)}$ until convergence |
| Label each point $x_i$ by $y_i = \operatorname{argmax}_{j \leq C} \hat{Y}_{ij}^{(\infty)}$ |

## 3. Experiments

### 3.1. Household simulation

We used the VoxCeleb1 [1] dataset to construct graphs and evaluate SID performance with different signals, LP and fusion methods. We evaluate SID performance in a simulated household scenario, reflecting the use case of most smart speaker AI assistants. To test model fairness regarding households with different demographic characteristics and speaker similarities, we utilized the speakers' gender and nationality metainformation provided by the dataset, and synthesized 6 groups of 4-speaker households:

- **Random**: Households with randomly selected speakers from VoxCeleb1, as in [9] (1,251 speakers in total).
- **Hard**: Households with hard-to-discriminate speakers as measured by speaker embedding distance. Speaker-level embedding profiles are calculated by averaging up to 100 utterances of the speaker. Then speakers with high similarity are selected based on 75th percentile among all speaker-to-speaker profile distances, similar to [8][12].
- **Same-gender**: Households with all speakers of the same gender (drawn from 690 male or 561 female speakers).
- **Same-locale**: Households with all speakers from the same locale (drawn from 799 USA or 215 UK speakers).

For each household group, we split the households into development and validation sets with a ratio of 1:2. The development set is used for optimizing hyperparameters while the validation set is used for final evaluation. For each household, 10 utterances per speaker are randomly selected to serve as the held-out dataset for evaluation. The rest of the utterances can be selected either as labeled (i.e., enrollment) or as unlabeled samples for the SSL experiments. We use the speaker identification error rate (SIER) within a household as the performance metric. SIER is defined as 1 – (accuracy of top predicted speaker). The final SIER is calculated as the micro-average over the households in the validation set.

### 3.2. Signals and embeddings

Multi-view fusion is applied to three kinds of signals:

- **Audio (voiceID)**: The speaker embedding extractor is trained on VoxCeleb2 [2] for the text-independent speaker verification using generalized end-to-end (GE2E) loss [20]. Here, voiceID embeddings with 512 dimensions are used as the main modality.
- **Visual (faceID)**: To obtain face embeddings corresponding to utterances, we use the LightFace [21], [22] package and its pretrained models. We first sample up to 10 images per utterance with a sampling rate of 1 image per second. We use MTCNN [23] for face extraction and alignment, and ArcFace [24] for generating the 512-dimensional faceID embeddings. The final utterance-level embeddings are generated by averaging over all the sampled images.
- **Meta (sessionID)**: in real smart speaker use cases, session defines a period of continuous customer engagement with the device. This metainformation can serve as a strong, but incomplete, constraint on speaker labels. For VoxCeleb1 data, we assume two utterances share the same sessionID only if they are from the same YouTube video. When constructing the graph with the view of sessionID, we assign a distance of 0 for two utterances sharing the same sessionID, and 1 otherwise.

### 3.3. Scoring methods

In real-time, each household comprises a different set of speakers. Thus, SID in a household scenario is usually treated as a speaker verification (SV) task, where the predicted label for each utterance is given by the household speaker with the highest SV score.

We evaluate LP-based approaches against four baselines:

- **CS**: Cosine scoring [25], [26]. We compute the cosine similarity score between each utterance in the held-out set and each speaker's labeled utterances only, and compute an average score per speaker. The speaker with the highest score is picked as the predicted label for a held-out utterance. No prior unlabeled utterances are used.
- **CSEA**: Cosine scoring with embedding average [20], [27]. For each speaker/class, we compute the speaker level representation by averaging across all embeddings from the labeled utterances. For each utterance in the held-out set, we compute the cosine similarity score to the speaker-level representation. The speaker with the highest score is the predicted label for a held-out utterance. No prior unlabeled utterances are used.
- **2-CS**: 2-step cosine scoring. In Step 1, we calculate pseudo-labels using the CS method for all unlabeled utterances. In Step 2, we predict the labels of utterances in the held-out set with the labeled utterances and pseudo-labels from Step 1, using the CS method.
- **2-CSEA**: 2-step cosine scoring with embedding average. In Step 1, we calculate pseudo-labels using the CSEA method for all unlabeled utterances. In Step 2, we predict the label for the held-out utterances with labeled utterances and pseudo-labels from Step 1, using CSEA.

We evaluated the following three LP-based methods:

- **LP**: Simple label propagation over a graph containing labeled, unlabeled and held-out samples for each household (Algorithm 1).
- **2-LP**: 2-step label propagation. In this method, Step 1 performs LP to compute predictions for each unlabeled datapoint, based on labeled datapoints. In Step 2, we utilize the predictions from Step 1 as pseudo-labels and perform a second round of LP to make final predictions for each utterance in the held-out dataset, using both labeled and unlabeled data.
- **2-LPEA**: 2-step method with label propagation as Step 1 and embedding average as Step 2. Step 1 performs LP to compute predictions for the unlabeled datapoints based on labeled data. In Step 2, we utilize the predictions from Step 1 as pseudo-labels and use the CSEA method to make final predictions for each utterance in the held-out dataset, using both labeled and unlabeled data.

The rationales for comparing these methods are: (1) CS and CSEA are the most commonly used methods for speaker verification in previous work, but they do not make use of unlabeled data. (2) 2-CS and 2-CSEA extend CS and CSEA by using traditional pseudo-labels on unlabeled data, giving us a suitable baseline for other SSL methods. (3) 2-LP extends LP by converting soft labels to hard labels for unlabeled samples and then normalizing the labels and pseudo-labels in the second step to have a more balanced label distribution for the classification of held-out data. (4) 2-LPEA is a practical approach with fixed runtime computation. Step 1 followed by embedding averaging computes a compact speaker representation offline, while benefitting from LP. Runtime processing involves only traditional scoring of embeddings.

## 3.4. Experimental results

We conducted evaluations by randomly selecting $L$ utterances per speaker and $U$ utterances per household as the labeled/unlabeled samples, respectively. The choice of $L$ and $U$ has been discussed in [9]. Here in this work, $L$=2 and $U$=320.

Table 1 summarizes single-view results for different groups of households and label prediction methods on the validation set. Only audio signals with voiceID embeddings are used. Four groups of methods are compared: (1) "Baselines": baseline methods as mentioned above; (2) "Universal scaling": LP-based methods with a universal scaling factor of $\sigma=0.22$, optimized for random households; (3) "Cohort scaling": LP-based methods with the cohort-specific scaling factor optimized for the target group of households, while being constant across households within a group; (4) "Local scaling": LP-based methods with the locally adapted scaling factor as per Equation (3), with $K$=40 and $s$=0.3, as optimized for random households. The numbers in bold refer to the best SIER within each method group. The relative improvements for each method group against the lowest SIER among the baselines are shown as well.

As Table 1 shows, though the universal scaling factor works for randomly assembled households, it is not optimal for other kinds of households, presumably due to demographic variations and different degrees of voice similarity (confusability). However, with locally adapted scaling factor, the 2-LP method consistently outperforms the baselines, universal and even cohort scaling methods, even though the hyperparameters ($K$ and $s$) are optimized with random households. Notably, even with a cohort-specific scaling factor, there still seems to be a majoritarian bias among individual households in that group. Thus, only the locally adapted graph normalization, which can capture individual household patterns, can effectively counteract this majoritarian bias and improve model robustness.

Table 2 summarizes results on "hard" households regarding different signals and fusion methods. Here, "V", "F" and "Session" represent the voiceID, faceID and sessionID signal, respectively. The best SIER for single-view signals are 2.41 and 2.78 for voiceID and faceID, respectively. However, by fusing them, both EdgePool and PML achieve substantial improvements over single-signal SID. Specifically, the PML ($p$=1) method achieves the best SIER, 0.53 for V+F fusion, an improvement of 75.2% over voice-only and 80.9% over face-only SID. Moreover, although the sessionID signal itself is not strong enough to support label prediction, it gives additional gains when fused with other signals. PML with different values of $p$ is also investigated, including cases referring to approximate max/min-pooling ($p$=5/-5), arithmetic ($p$=1), and harmonic ($p$=-1) mean of Laplacian matrices [11].

It is noteworthy that 2-LP produces the best performance for almost all cases, except for those based on faceID and not using PML. One interpretation is that, with CSEA as the last step, 2-LPEA is more robust to signals with embedding outliers, while 2-LP is more accurate with clean data. Error analysis showed that most of the mistakes involving the faceID signal are from failure to detect the target speaker in images with multiple faces, while working accurately elsewhere. Thus, the faceID signal is a strong signal, but has outliers. In principle, 2-LPEA works better with faceID signal. However, PML is a powerful fusion method that can blend the information encoded across multiple views, and is robust to noise [11]. As a result, 2-LP with PML achieves the best performance for all fusion tasks. We conclude that multi-view graph-SSL is effective at fusing even noisy or incomplete signals.

Table 1: *SIER (%) on validation set with voiceID embeddings* ($L$=2, $U$=320). Best results boldfaced.

| Method | | Voice similarity | | Same-gender | | Same-locale | |
|---|---|---|---|---|---|---|---|
| | | random | hard | male | female | USA | UK |
| Baselines | CS | 3.36 | 8.85 | 5.37 | 5.88 | 3.76 | 4.68 |
| | CSEA | 3.06 | 7.88 | 4.69 | 5.42 | 3.63 | 3.93 |
| | 2-CS | 1.71 | 4.07 | 2.24 | 2.37 | 1.96 | 1.53 |
| | 2-CSEA | **1.44** | **3.46** | **1.94** | **2.10** | **1.59** | **1.38** |
| Universal scaling | LP | 1.46 | 3.79 | 2.31 | 2.80 | 1.76 | 1.85 |
| | 2-LP | 1.37 | 3.27 | 2.02 | 2.61 | **1.65** | 1.79 |
| | 2-LPEA | **1.31** | **3.22** | **1.86** | 2.25 | 1.69 | **1.68** |
| Cohort scaling | LP | 1.46 | 4.24 | 2.44 | 3.66 | 1.69 | 2.19 |
| | 2-LP | 1.37 | 3.93 | 2.33 | 3.52 | **1.59** | 2.04 |
| | 2-LPEA | **1.31** | **3.14** | **1.74** | **1.99** | 1.68 | **1.41** |
| Local scaling | LP | 1.28 | 3.24 | 1.92 | 2.06 | 1.66 | 2.31 |
| | 2-LP | **0.92** | **2.41** | **1.72** | **1.58** | **1.50** | **1.29** |
| | 2-LPEA | 1.19 | 2.99 | 1.74 | 1.82 | 1.70 | 1.56 |
| Improvement (%) | Universal | 9.0 | 6.9 | 4.1 | -7.1 | -3.8 | -21.7 |
| | Cohort | 9.0 | 9.2 | 10.3 | 5.2 | 0.0 | -2.2 |
| | Local | **36.1** | **30.4** | **11.3** | **24.8** | **5.7** | **6.5** |

Table 2: *SIER (%) on hard households using multiple signals and fusion methods* ($L$=2, $U$=320).

| Signal (Fusion) | Baselines | | | | Graph-based | | |
|---|---|---|---|---|---|---|---|
| | CS | CSEA | 2-CS | 2-CSEA | LP | 2-LP | 2-LPEA |
| Voice (V) only | 8.85 | 7.88 | 4.07 | 3.46 | 3.24 | **2.41** | 2.99 |
| Face (F) only | 3.57 | 3.52 | 2.95 | 2.95 | 3.43 | 3.02 | **2.78** |
| V+F (EdgePool) | - | - | - | - | 3.16 | 2.76 | **2.03** |
| V+Session (EdgePool) | - | - | - | - | 2.58 | **2.33** | 2.72 |
| V+F+Session (EdgePool) | - | - | - | - | 2.51 | 2.22 | **1.93** |
| V+F (PML, $p$=1) | - | - | - | - | 0.75 | **0.53** | 1.76 |
| V+Session (PML, $p$=1) | - | - | - | - | 2.18 | **1.83** | 2.42 |
| V+F+Session (PML, $p$=1) | - | - | - | - | 0.66 | **0.54** | 1.79 |
| V+F+Session (PML, $p$=5) | - | - | - | - | 2.02 | **1.49** | 2.17 |
| V+F+Session (PML, $p$=2) | - | - | - | - | 0.94 | **0.74** | 1.81 |
| V+F+Session (PML, $p$=-1) | - | - | - | - | 0.78 | **0.54** | 1.78 |
| V+F+Session (PML, $p$=-2) | - | - | - | - | 0.82 | **0.59** | 1.78 |
| V+F+Session (PML, $p$=-5) | - | - | - | - | 0.97 | **0.75** | 1.80 |

## 4. Conclusions

We have proposed a graph-based semi-supervised inference approach for improving speaker identification in household scenarios. The proposed approach improves household-level SID by leveraging (1) both labeled and unlabeled data with SSL; (2) household-specific and local information with locally adapted graph normalization; (3) benefits from multi-modal and metadata signals with multi-view fusion of graphs. We evaluated the method, in several variants, on VoxCeleb data, using simulated households with different demographic characteristics and levels of within-household voice similarity. The experiments with baseline methods, as well as LP-based ablation experiments, demonstrate that the proposed approaches are effective at improving SID accuracy and robustness to different household properties. Furthermore, this approach provides a way to carry out household-specific adaptation and multi-signal fusion without retraining embeddings or a fusion network.

## 5. Acknowledgments

We would like to thank the Alexa Speaker Understanding team for their valuable input, and Mohamed El-Geish for supporting this work.